\documentclass[letter]{aa}

\usepackage{graphicx}
\usepackage{txfonts}
\usepackage{natbib}

\newcommand{\kms}{km~s$^{-1}$}
\newcommand{\dg}{$^{\circ}$}
\newcommand{\rsun}{$R_{\odot}$}

\newcommand{\Tii}{Type~{{{\sc{ii}}}}}
\newcommand{\tii}{type~{{{\sc{ii}}}}}
\newcommand{\tiii}{type~{{{\sc{iii}}}}}

\begin{document}

\title{Radio evidence for a shock wave reflected by a coronal hole}
\author{S. Mancuso\inst{1}, A. Bemporad\inst{1}, F. Frassati\inst{1}, D. Barghini\inst{1,2}, S. Giordano\inst{1}, D. Telloni\inst{1}, \and C. Taricco\inst{1,2}}

\institute{
${^1}$ Istituto Nazionale di Astrofisica, Osservatorio Astrofisico di Torino, via Osservatorio 20, Pino Torinese 10025, Italy \\ 
${^2}$  Università degli Studi di Torino~--~Dipartimento di Fisica, via Pietro Giuria 1, Torino (TO), Italy \\ \email{salvatore.mancuso@inaf.it}}
\date{Received / Accepted}

\abstract{
We report the first unambiguous observational evidence in the radio range of the reflection of a coronal shock wave at the boundary of a coronal hole.
The event occurred above an active region located at the northwest limb of the Sun and was characterized by an eruptive prominence and an extreme-ultraviolet (EUV) wave steepening into a shock.
The EUV observations acquired by the Atmospheric Imaging Assembly (AIA) instrument on board the {\it Solar Dynamics Observatory} ({\it SDO}) and the Extreme Ultraviolet Imager (EUVI) instrument on board the {\it Solar TErrestrial RElations Observatory} ({\it STEREO-A}) were used to track the development of the EUV front in the inner corona.
Metric \tii\ radio emission, a distinguishing feature of shock waves propagating in the inner corona, was simultaneously recorded  by ground-based radio spectrometers.
The radio dynamic spectra displayed an unusual reversal of the \tii\ emission lanes, together with \tiii-like herringbone emission, indicating shock-accelerated electron beams.
Combined analysis of imaging data from the two space-based EUV instruments and the Nan\c{c}ay Radioheliograph (NRH) evidences that the reverse-drifting \tii\ emission was produced at the intersection of the shock front, reflected at a coronal hole boundary, with an intervening low-Alfvén-speed region characterized by an open field configuration. 
We also provide an outstanding data-driven reconstruction of the spatiotemporal evolution in the inner corona of the shock-accelerated electron beams produced by the reflected shock. 

}
\titlerunning{Radio evidence for a shock wave reflected by a coronal hole}
\authorrunning{Mancuso et al.}
\maketitle

\section{Introduction}

\Tii\ radio bursts in the solar corona are usually characterized by a pair of slowly drifting bands of enhanced radio emission decreasing from high to low frequencies in radio dynamic spectra.
The observed frequency drift is usually attributed to the outward propagation of a shock wave, caused by either a flare or a coronal mass ejection (CME), away from the Sun.
Only in extremely rare cases have \tii\ radio bursts with reverse (positive) frequency drift been investigated (e.g., \citealt{Markeev1976,Korolev1979,Mancuso2004,Kumar2016}).
Within the framework of the plasma emission hypothesis, the generation of \tii\ radio bursts takes place at frequencies close to the Langmuir frequency of the plasma and its second harmonic.
Assuming generation of \tii\ emission by shock waves, the observed reverse frequency drift would imply shock propagation toward the solar surface or at some angle in the direction of a local enhancement of the coronal electron density. 
However, since \tii\ radio bursts with reverse frequency drift in the metric band are uncommon, a likewise extraordinary physical mechanism must be at work to account for their production.

Coronal shock waves are often, but not always, associated with extreme ultraviolet (EUV) waves.
These coronal disturbances are usually seen as moving hemispherical disturbances in EUV emission traveling away from their source regions with speeds ranging from $50-700$ \kms\ (e.g., \citealt{Thompson1998,Warmuth2011,Nitta2013,Shen2014,Long2017}).
Several possible interpretations of these globally propagating EUV disturbances have been proposed in the literature, such as magnetohydrodynamic (MHD) fast- and slow-mode wave models, pseudo-wave models, and hybrid models involving both scenarios (see the review of \citealt{Warmuth2015}).
A long-term debate over these competing interpretations still exists.
However, the wave interpretation is supported, at least in some cases, by the observation of reflections and refractions at regions with strong gradients in Alfvén speeds, such as at the boundaries of active regions and coronal holes.
Most notably, reflections of EUV waves at coronal hole boundaries have been reported by several authors (e.g., \citealt{Long2008,Gopalswamy2009,Li2012,Olmedo2012,Shen2013,Kienreich2013,Kumar2013}).
These detections have recently been supported by numerical simulations (\citealt{Piantschitsch2017,Afanasyev2018,Xie2019}).
Extreme-ultraviolet waves can eventually evolve into shocks when their speeds exceed the magnetosonic speed of the ambient plasma, which can be evidenced by the appearance of \tii\ radio bursts when favorable conditions are satisfied, such as quasi-perpendicularity or interaction with adjacent low-Alfvén-speed coronal structures along the expanding flanks (e.g., \citealt{Mancuso2019}). 
Reflection of shock waves in the inner corona is certainly expected as well. 
However, surprisingly enough, incontrovertible direct observational evidence in the radio band is still lacking.

In this letter we analyze a solar eruption characterized by a coronal EUV front steepening into a shock wave and report, for the first time, direct observational evidence in the radio band of the sunward reflection of a portion of the coronal shock wave surface at the boundary of a coronal hole.
The sunward propagation of the reflected shock wave has been revealed in an unprecedented way by means of radioheliograph observations of the electron beams accelerated at the shock in correspondence with a \tii\ radio burst with reverse frequency drift.

\begin{figure}
    \centering
    \includegraphics[width=7.5cm]{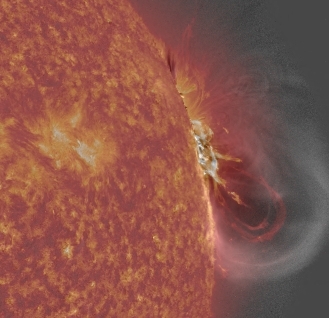}
    \caption{Prominence eruption of 2011 August 11 as seen by {\it SDO}/AIA in the 304 \AA\ channel at 10:12 UT.
    A superposed gray layer shows a running-difference image, taken in the 211 \AA\ channel, outlining the location of the EUV wave front.}
    \label{Fig01}
\end{figure}

\begin{figure}[b]
\centering
\includegraphics[width=8.4cm,trim=0cm 7cm 0cm 0cm, clip=true]{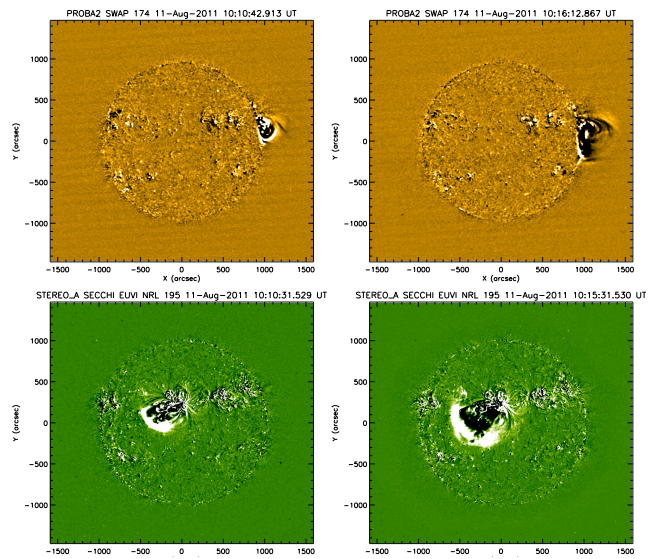}
\caption{Evolution of the CME as seen from the perspective of {\it PROBA2}/SWAP (top panels) and {\it STEREO-A}/EUVI (bottom panels) at three different times.}
\label{Fig02}
\end{figure}

\begin{figure*}
\centering
\includegraphics[width=15cm]{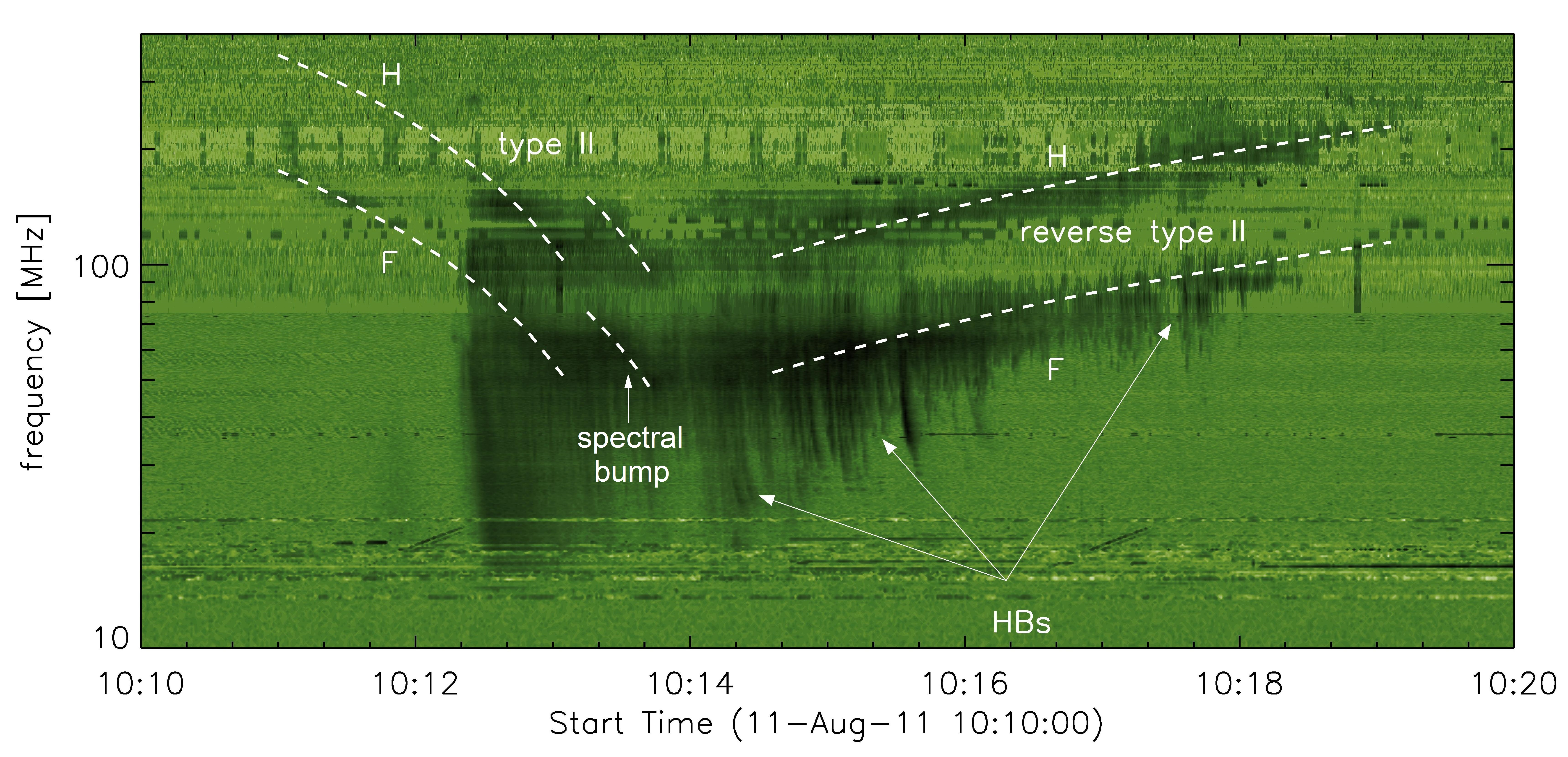}
\caption{
Combined radio dynamic spectrum of NDA (10 -- 80 MHz) and CALLISTO (80 -- 400 MHz). 
Dashed white lines outline the F and H lanes of the observed \tii\ bursts. 
The first two F and H emission lanes (with frequency decreasing with time) are connected to the shock ahead of the CME eruption, while the last two lanes (with frequency increasing with time) represent the BBs of the \tii-like emission attributed to the reflected shock.
}
\label{Fig03}
\end{figure*}

\section{Observations and data reduction}

The 2011 August 11 eruption originated from NOAA Active Region (AR) 11263, located just behind the western limb of the visible solar disk as seen from Earth.
The solar eruption produced a C6.2 class soft X-ray flare on the Geostationary Operational Environment Satellite (GOES) scale and a partial fast halo CME that was detected in the higher corona on the west limb of the Sun by the Large Angle and Spectroscopic Coronagraph (LASCO C2) on board the {\it Solar and Heliospheric Observatory} ({\it SOHO}).
According to the online CACTUS Database\footnote{\tt http://sidc.oma.be/cactus/catalog.php}, the CME propagated at a median speed of $976 \pm 309$ \kms, with a maximum speed of 1602 \kms. 
A spectacular large inclined filament eruption followed the flare in AR 11263, and the ejected prominence material was seen traveling southward in the 304 \AA\ channel (see Fig.~\ref{Fig01}), dominated by the He {\sc{ii}} line ($\log T [\rm K]\sim 4.7$), of the Atmospheric Imaging Assembly (AIA) instrument on board the {\it Solar Dynamics Observatory} ({\it SDO}).
An EUV wave propagated ahead of the prominence eruption, as evinced by running-difference images in the 211 \AA\ channel dominated by the Fe {\sc{xiv}} line ($\log T [\rm K] \sim 6.3$).
Figure~\ref{Fig02} displays the evolution of the EUV wave at different times from Earth's perspective, as seen in the 174 \AA\ passband (Fe {\sc{ix}/\sc{x}}; $\log T [\rm K]\sim 6.0$) by the Sun-Watcher with Active Pixel System and Image Processing (SWAP) instrument on board the {\it Project for On-Board Autonomy 2} ({\it PROBA2}) and in the 195 \AA\ channel (Fe {\sc{xii}}; $\log T [\rm K] \sim 6.2$) by the Extreme Ultraviolet Imager (EUVI) instrument on board the {\it Solar TErrestrial RElations Observatory} ({\it STEREO-A}), separated from Earth by 100.9\dg. 
At the time of the eruption, AR 11263 was near central meridian from the perspective of {\it STEREO-A}, which was thus favorably located.

\begin{figure*}[h]
\centering
\includegraphics[width=7.4cm]{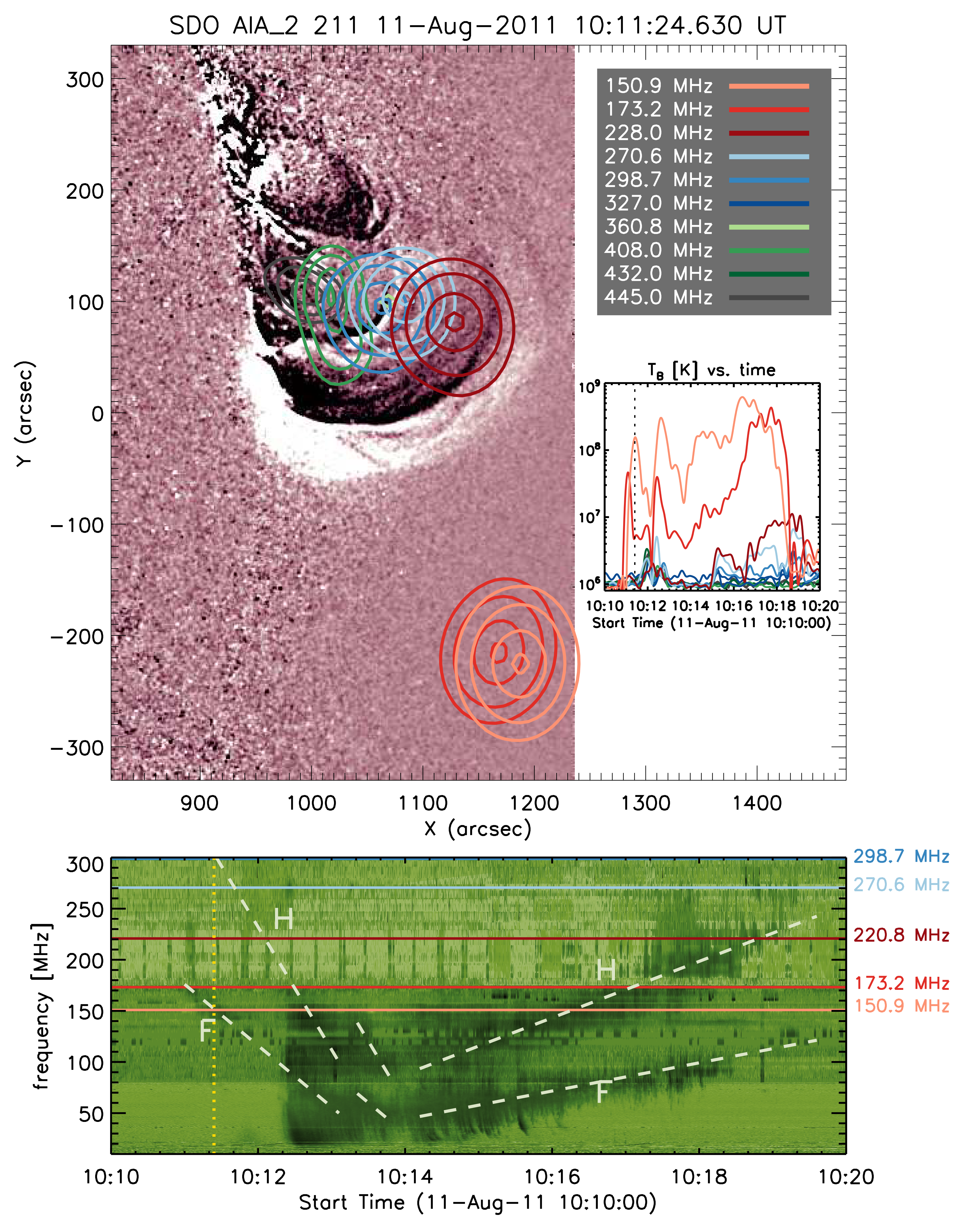}
\includegraphics[width=7.4cm]{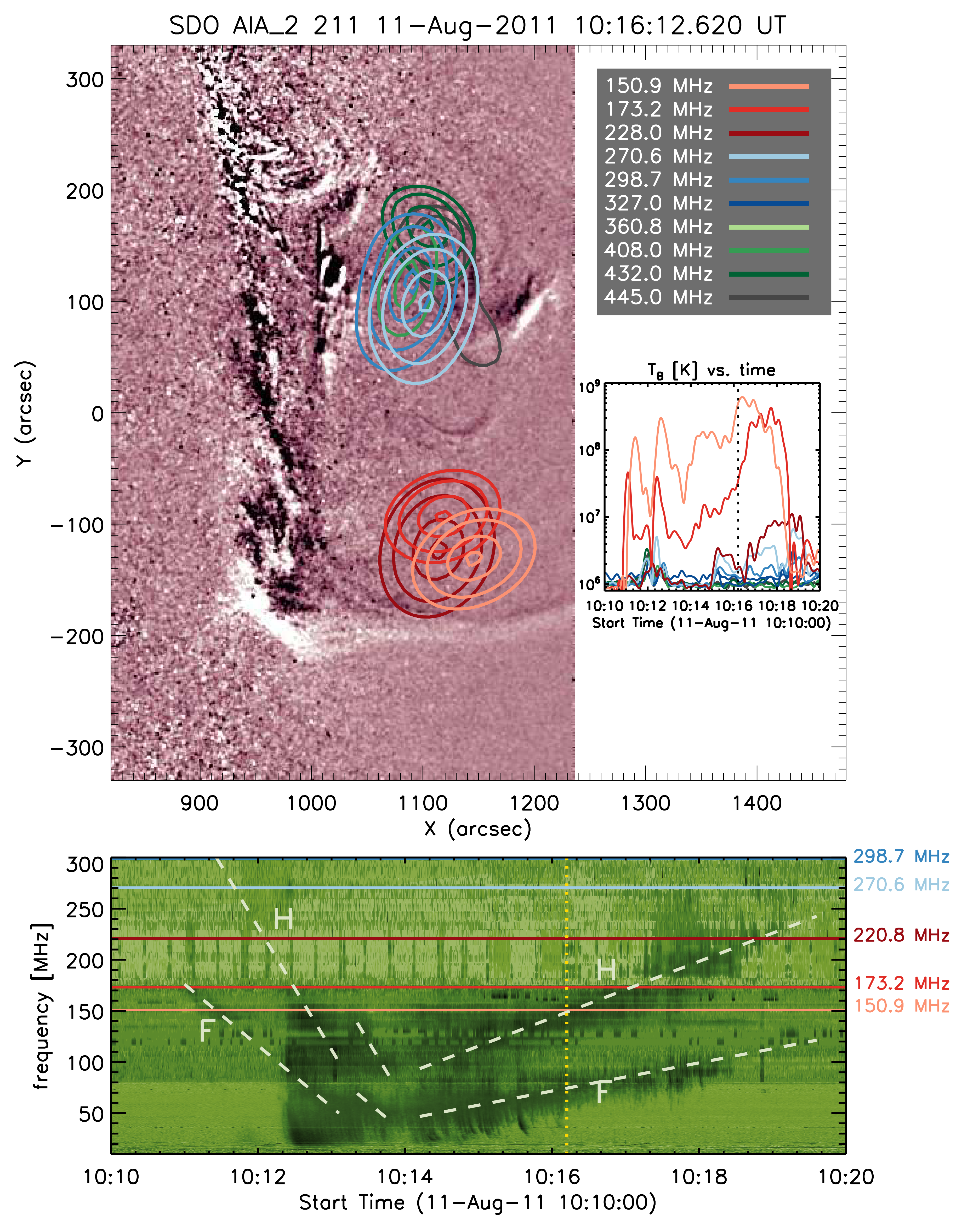}
\caption{Contour plots of the locations of the \tii\ radio emission with respect to the eruption as seen by simultaneous images taken by AIA in the 211 \AA\ channels. 
Contour levels represent 95\%, 97\%, 99\%, and 99.9\% of the maximum temperature brightness of the full NRH radio map.
The positively drifting \tii\ emission ({\it left panel}) seen at frequencies of 150.9 and 173.2 MHz at the F was emitted around 10:11.5 UT and is clearly well ahead of the CME eruption. 
The right panel shows the position of the negatively drifting \tii\ emission at a later time.
The simultaneous radio emissions at frequencies higher than 200 MHz are not related to the shock but are rather attributable to hot prominence material expanding during the eruption. }
\label{Fig04}
\end{figure*}

\begin{figure*}
\centering
\includegraphics[width=18cm]{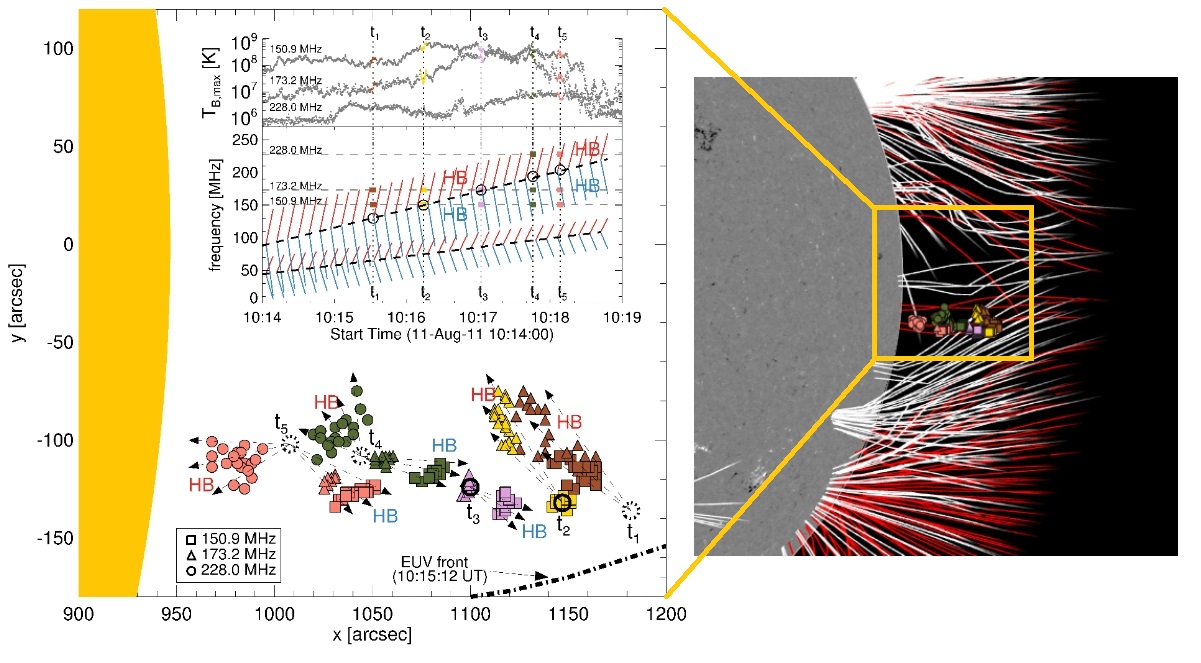}
\caption{
{\it Left panel:} 
Spatiotemporal evolution of the BB + HB emission at 150.9, 173.2, and 228.0 MHz.
Different symbols and colors specify, respectively, the frequencies and times $t_k$ at which the HBs were observed.
HBs have been imaged at the H frequency:
Only the position of the maximum temperature brightness ($T_{B,\rm max}$) is depicted.
Solid circles at times $t_2$ and $t_3$ indicate the observed positions of the BB emission; dotted circles at times $t_1$, $t_4$, and $t_5$ denote inferred positions of the BB emission on the basis of the observed HB emissions.
Dashed arrows departing from solid and dotted circles identify the directions of the electron beams producing the observed HB emission at each time $t_k$.
The bottom panel of the inset shows a simplified overview of the observed HB structure in the time interval 10:14--10:19 UT, while the upper panel shows the maximum $T_B$ at each frequency.
{\it Right panel:} Coronal magnetic field reconstructed with the PFSS technique up to 1.25 \rsun.
White lines correspond to closed magnetic field loops, red lines to open fields. 
The observed radio sources are aligned along open field lines. 
The plot was obtained with the JHelioviewer tool (\citealt{Muller2017}).
}
\label{Fig05}
\end{figure*}

The EUV wave steepened into a shock around 10:11 UT, as evidenced by the appearance of a metric \tii\ radio burst in ground-based radio spectrometers.
As a side note, we point out that the shock may have actually been ignited somewhat before since \tii\ emission, as previously discussed, requires specific favorable conditions to be satisfied.
The radio dynamic spectrum displayed in Fig.~\ref{Fig03} is a composite obtained by combining data from two spectrometers belonging to the e-Callisto ($45-870$ MHz) network\footnote{\tt http://www.e-callisto.org} with those from the Nançay Decameter Array (NDA; $10–100$ MHz) solar radio spectrograph. 
\Tii\ radio bursts often reveal a variety of morphological features, such as a backbone (BB), fundamental (F) and harmonic (H) emissions, herringbone (HB) structure, etc.
The BB emission corresponds to the slowly drifting emission lane considered as the signature of a shock wave traveling away from the Sun through the corona.
Although the data are heavily affected by the presence of interference, especially between about 80 and 110 MHz, the patchy, slowly negatively drifting F-H structure of the \tii\ burst, starting at about 10:11 UT, is evinced from the composite radio dynamic spectra. 
A spectral bump, probably caused by the interaction of the shock with a denser coronal structure (e.g., \citealt{Feng2013}), is also visible in Fig.~\ref{Fig03} between 10:13 and 10:14 UT.
The position of the radio sources corresponding to the emission from the negatively drifting BB H lane (as imaged around 10:11 and 10:12 UT) was found to be just ahead of the EUV front, as expected in the case of a CME-driven shock enveloping the outwardly expanding eruption (see Fig.~\ref{Fig04}).
The observations were carried out in the metric band by the Nançay Radioheliograph (NRH), which, at the time of the event, was observing the Sun with high temporal cadence (0.25 s).
\Tii\ radio bursts occasionally exhibit a \tiii-like fine HB structure emanating from the BBs toward both high and low frequencies with typical drift rates of about $\pm 10$ MHz s$^{-1}$.
In Fig.~\ref{Fig03}, both F and H HBs can be seen throughout the entire radio event.
Their presence is generally attributed to plasma emission from electron beams accelerated by shock drift acceleration (SDA), which is particularly efficient in the case of quasi-perpendicular propagation (\citealt{Holman1983}).
Although SDA only produces electron beams in the upstream region, electrons accelerated by a curved shock can propagate upstream along open magnetic field lines toward both the higher and lower corona.
In this case, the presence of HBs throughout the event thus suggests an open field configuration and quasi-perpendicular shock propagation (see also \citealt{Carley2013,Carley2015,Morosan2019}).

\begin{figure}[t]
\centering
\includegraphics[width=8.6cm]{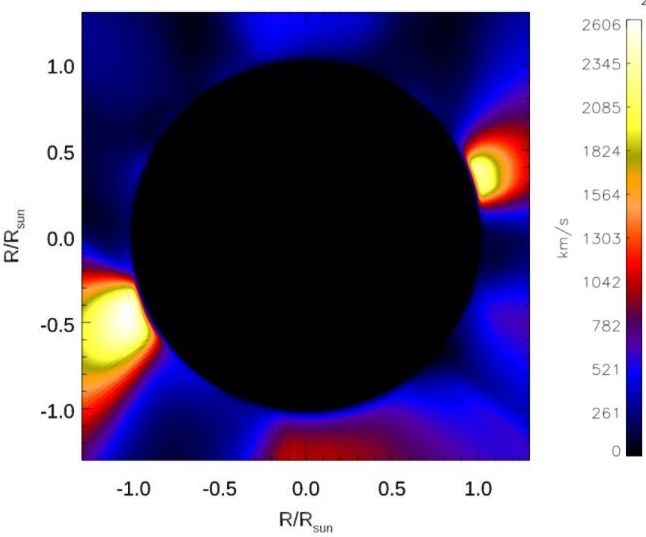}
\caption{
Coronal Alfvén speed on 2011 August 11 at 9:00 UT as obtained by Predictive Science Inc. in the context of the MAS model (\tt http://www.predsci.com/mhdweb/home.php). 
}
\label{Fig06}
\end{figure}

\begin{figure*}
\centering
\includegraphics[width=5.0cm]{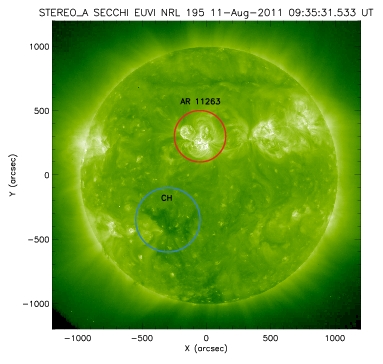}
\includegraphics[width=12.8cm]{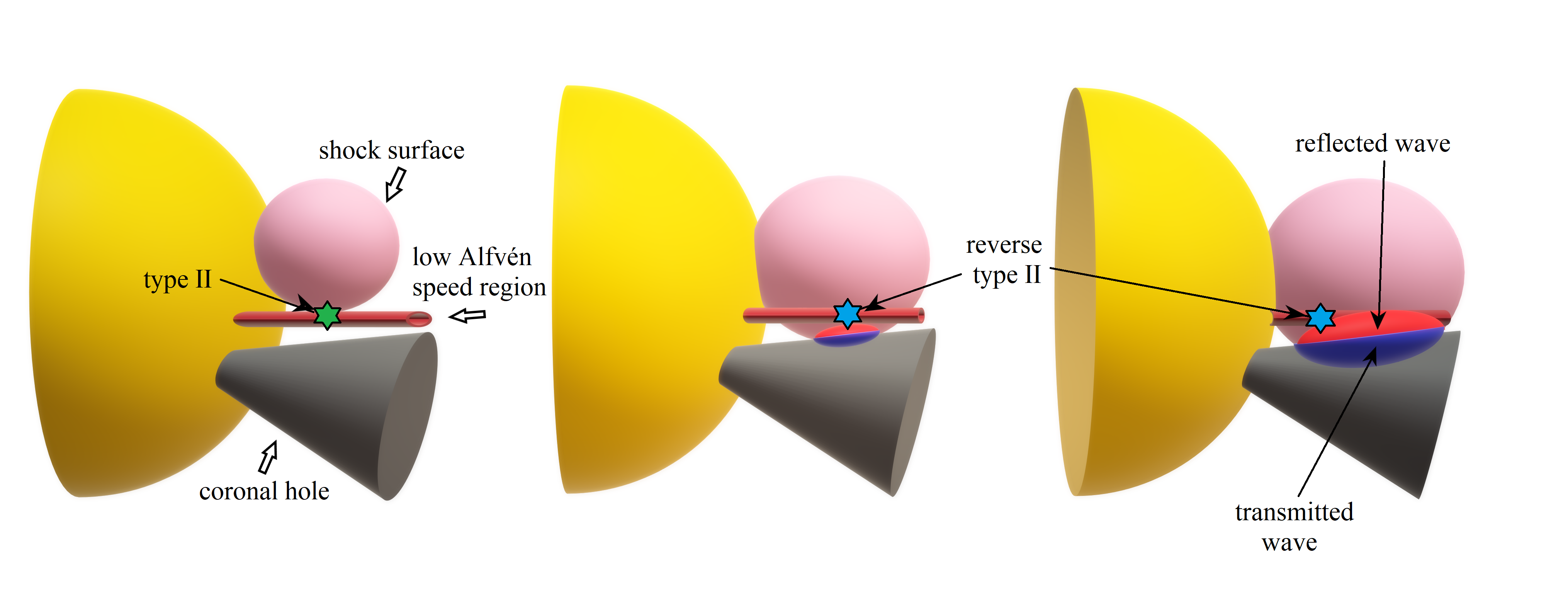}
\caption{
{\it Left panel:} {\it STEREO-A}/EUVI 195 \AA\ image of the corona before the eruption. A coronal hole (CH) is clearly visible in the southern hemisphere.
{\it Right panel:} Cartoon showing the proposed geometry involved in the \tii\ emission at three times during the shock evolution. The \tii\ emission is attributed to the interaction of the transmitted (green star) and reflected (light blue star) shock surface with a narrow low-Alfvén-speed region near the equator.}
\label{Fig07}
\end{figure*}

\section{Discussion and conclusions}

The peculiarity of this complex radio burst is the unusual reversal of the F and H BBs toward higher frequencies occurring between about 10:14 and 10:19 UT.
As we have already mentioned, this is an extraordinary feature that has hardly ever been observed in radio dynamic spectra and, to our knowledge, never imaged by radio heliographs.
As depicted in the left panel of Fig.~\ref{Fig05}, the three lowest NRH frequencies (150.9, 173.2, and 228.0 MHz) luckily covered the H band of the BB + HBs reverse \tii\ emission, so we were able to discern and distinguish their actual positions (see also Fig.~\ref{Fig04}).
In Fig.~\ref{Fig05} we also reconstruct the spatiotemporal evolution of the BB + HBs reverse \tii\ emission as observed at the five times specified in the inset (shown in different colors). 
The locations of the centroids of the radio sources were estimated by determining, for each 2D radio intensity map, the maximum brightness observed at each frequency after smoothing the radio maps with a Gaussian kernel using the {\tt gauss\_smooth.pro} Interactive Data Language (IDL) procedure. 
The maps were then re-binned to a finer grid to obtain the positions of the centroids of the radio sources in the image plane at each time and at each given frequency.
In Fig.~\ref{Fig05}, solid circles at times $t_2$ and $t_3$ yield the actual position in the plane of the sky of the BB radio source (the shock-related emission), while dotted circles yield a qualitative indication of the possible position of the BB at times $t_1$, $t_4$, and $t_5$ according to the observed positions of the HB sources.
A simplified overview of the observed BB + HBs structure in the time interval is also shown in the same figure to facilitate the reader's overview of the event.
The shock-related radio emission from the H BB of the reversely drifting \tii\ radio burst is clearly seen to propagate radially toward the Sun.

To identify the mechanism responsible for the observed anomalous \tii\ reverse drift, it is imperative to determine the underlying magnetic field configuration.
In the right panel of Fig.~\ref{Fig05}, we display the coronal magnetic field up to 1.25 \rsun\ as derived from a potential field source surface (PFSS; \citealt{Schrijver2003}) model based on Helioseismic and Magnetic Imager ({\it SDO}/HMI; \citealt{Schou2012}) magnetograms: White lines correspond to closed magnetic field loops, while red lines denote open fields.
Superposed on the image, we also show for comparison the radio emission from the reverse \tii\ radio feature previously described.
From the extrapolated field, we infer that the BB + HBs radio emission was emitted in correspondence with an open field region, thus implying that the accelerated particles could easily escape the shock front via SDA.
The appearance of \tii\ radio sources can be thought of as a sort of visualization of the low-Alfvén-speed structures existing in the corona, which literally lightens up from the shock that strengthens in that region (e.g., \citealt{Uchida1974}), with the radio emission also being favored by the quasi-perpendicularity of the field.
In Fig.~\ref{Fig06} we show the Alfvén speed distribution in the plane of the sky as calculated at 9:00 UT by Predictive Science Inc. in the context of the Magnetohydrodynamic Around a Sphere (MAS) model. 
The image clearly shows, as expected, a low-Alfvén-speed region in the region under investigation.
On the other hand, since we just have a 2D representation of the radio emission, this evidence is only circumstantial. 
It is clear, however, given the geometry of the observed EUV wave (see Fig.~\ref{Fig01}) and the field configuration, that the enveloping shock surface propagated almost perpendicularly to the upstream field.
In principle, the sunward directed \tii\ emission (and the related reversely orientated spectral signature) could be excited by the expanding shock surface while traversing and intersecting the abovementioned low-Alfvén-speed region, as already proposed by \cite{Mancuso2004}.
However, at the time of the reverse \tii\ emission, the EUV wave (and even more so the shock surface, apart from possible projection effects along the line of sight) was positioned much farther than the location where the radio emission was excited (see Fig.~\ref{Fig04}).
An alternative explanation is thus required.

The positively drifting \tii\ structures have also been interpreted as being due to downward particle acceleration at 
a termination shock generated by a reconnection outflow (\citealt{Aurass2002,Mann2009,Chen2015}) or to slippage of 
field lines causing the shift in the reconnection point (\citealt{Kumar2016}). 
However, the event reported in this letter occurred very far from the flaring active region and was produced 
in a region where,  according to the field extrapolation, no oppositely directed magnetic field lines were observed; as such, we can rule out the above hypotheses.
\cite{Piantschitsch2017} used a 2.5D numerical code to perform simulations of fast-mode MHD wave propagation in the corona and its interaction with coronal holes.
In their simulations, they presented the temporal evolution of the incoming wave, its impact with a low-density region characterized by high Alfvén speed, as in coronal holes, and the subsequently evolving secondary reflected, transmitted, and traversing waves.
As the wave moved toward the coronal hole, they observed a steepening of the wave that subsequently developed into a shock. 
At the coronal hole boundary, a reflection of the incident wave occurred as an immediate result of the impact of the wave on the coronal hole.
As a matter of fact, a large coronal hole was visible in the {\it STEREO-A}/EUVI images at lower heliolatitudes (see Fig.~\ref{Fig07}), and a clear interaction between the EUV wave and the coronal hole boundary could be seen in the 195 \AA\ passband (see Fig.~\ref{Fig02}), hinting that the shock reflection must have occurred, as expected, somewhat earlier than 10:15 UT.
Our proposed mechanism is exemplified in the cartoon shown in Fig.~\ref{Fig07}: The reverse \tii\ radio spectral feature was emitted at the intersection of the shock wave, reflected at the coronal hole boundary, with an intervening low-Alfvén-speed region characterized by an open field configuration.

\begin{acknowledgements} 
We thank the referee for comments that helped to improve
the paper and the teams of e-Callisto, NDA, NRH, SDO/AIA and Predictive Science Inc. for their open-data use policy. 
F. F. is supported through the Metis programme funded by the Italian Space Agency (ASI) under the contracts to the co-financing National Institute of Astrophysics (INAF): Accordo ASI-INAF n. 2018-30-HH.0
\end{acknowledgements}

\bibliographystyle{aa}
\bibliography{biblio}

\end{document}